\documentclass[aps,prb,twocolumn,superscriptaddress]{revtex4}

\usepackage{graphicx}
\usepackage{amsfonts}
\usepackage{amssymb}
\usepackage[]{amsmath}
\usepackage[]{latexsym}
\usepackage[]{float}
\usepackage{bm}
\newcommand{\mvecfour}[4]
{
\left(\begin{array}{c}
#1  \\
#2  \\
#3  \\
#4  
\end{array}
\right)
}

\begin{document}

% number in the upper right-hand corner of the title page in preprint mode.
%\preprint{}

\title{Topologically protected Landau levels in bilayer graphene in finite electric fields}

\author{Tohru Kawarabayashi}
%\thanks{}
\affiliation{Department of Physics, Toho University,
Funabashi, 274-8510 Japan}

\author{Yasuhiro Hatsugai}
%\thanks{}
\affiliation{Institute of Physics, University of Tsukuba, Tsukuba, 305-8571 Japan}
\affiliation{Kavli Institute for Theoretical Physics, University of California,
Santa Barbara CA 93106, USA}

\author{Hideo Aoki}
%\thanks{}%\altaffiliation{}
\affiliation{Department of Physics, University of Tokyo, Hongo, 
Tokyo 113-0033 Japan }

\date{\today}

\begin{abstract}
The zero-energy Landau level of bilayer graphene is shown to be 
anomalously sharp (delta-function like) against bond disorder  
as long as the disorder is correlated over a few lattice constants.
The robustness of the zero-mode anomaly 
can be attributed to the preserved chiral symmetry. 
Unexpectedly, even when we apply a finite potential difference 
(i.e., an electric field) between the 
top and the bottom layers, the valley-split $n=0$ Landau levels remain 
anomalously sharp although they are now 
shifted away from the zero energy, while the $n=1$ Landau levels 
exhibit the usual behavior.  
\end{abstract}

%\keywords{}

\maketitle

{\it Introduction ---} 
The existence of  the zero-energy Landau level is a most fundamental property of the electronic states in
graphene in magnetic fields, which hallmarks the 
unconventional quantum Hall effect observed in  monolayer graphene. \cite{Geim,Kim}
Specifically, the zero-energy Landau level of mono-layer graphene  shows an anomalous robustness 
against the disorder induced by ripples, an intrinsic disorder in graphene, which leads to 
an unconventional criticality of Hall transition at zero energy.\cite{OGM,Guinea}
For the robustness of zero modes, 
the chiral symmetry,\cite{LFSG} defined in terms of the 
chiral operator $\Gamma$ that anti-commutes with the Hamiltonian $H$, $\{ \Gamma , H\}=0$ with 
$\Gamma^2=1$, is an essential ingredient.\cite{Hatsu10}
For monolayer graphene, we have vertical Dirac cones at K and K' points in 
the Brillouin zone, and  the effective Hamiltonian has the  
chiral symmetry.  In such a system, it has been demonstrated 
by the present authors that the zero-energy $(n=0)$ Landau level is robust  
against the disorder that respects the chiral symmetry as 
long as the disorder is correlated over a few lattice constants.\cite{KHA,KMHA}  
Experimentally, the $n=0$ Landau level narrower than the other $n\neq 0$ Landau levels is reported for monolayer graphene,\cite{GZKPMM}
which is consistent with the present robustness specific to the zero-energy ($n=0$) Landau level. 

The notion of the chiral symmetry is so universal 
that  it has further been shown \cite{KHMA}
that the chiral symmetry, usually 
considered for the vertical Dirac cones, can be generalized to 
accommodate 
tilted Dirac cones, such as those 
encountered in certain organic metals.\cite{KKS,GFMP,TSKNK} 
The generalized 
chiral symmetry protects the zero-energy Landau level 
as far as the Hamiltonian as a differential operator is elliptic, 
where we can even extend the argument of 
Aharonov and Casher for counting the number of 
zero modes in the presence of disorder.\cite{AC}
The existence of the 
generalized chiral symmetry can indeed be translated 
to a condition that  
the index theorem\cite{Nakahara} holds for generic tilted Dirac cones. 
The chiral symmetry is therefore directly related to the robustness of zero modes for the massless Dirac fermions.

Now, in the physics of graphene, 
the case of bilayer graphene is an interesting test bench 
for examining various graphene properties.  
Specifically, McCann and Falko have shown 
that there exist four-fold degenerated (per spin) zero-energy Landau levels, which lead to a 
quantum Hall effect characteristic to bilayer graphene.\cite{MF} 
The degeneracy comes from the valley (K and K') degrees of freedom and two ($n=0$ and $n=1$) Landau indices.  
Although the robustness of these zero mode Landau levels is also predicted 
as a consequence of the index theorem,\cite{KP,Kail} 
it is not clear whether there is also a direct relationship between the 
{\it chiral symmetry} and 
the anomalous robustness of zero modes, since 
the parabolic band dispersion in the bilayer graphene, 
as opposed to the linear one in the monolayer graphene, 
might well invalidate the arguments.  
For instance, the robustness of the $n=0$ level and that of $n=1$ level can naively be 
different, since they have different structure for the wave functions.

The difference becomes even greater when we 
apply an electric field perpendicular to 
the graphene sheet, which introduces an energy gap.  
An opening  of the energy-gap in bilayer systems is 
important in an applicational context as well.\cite{OBSHR,Castro,Neto} 
Thus a further interest is to see what happens to the zero-energy Landau level when 
the energy gap is introduced.  
Experimentally, it is desirable to clarify quantitatively the robustness 
of the zero modes in bilayer graphene, since 
experimental results in high mobility 
samples are now available.\cite{Weitz,Martin}

The purpose of the present paper is to explore these very questions, 
for which we have performed numerical studies based on the 
lattice model.  We shall show that both the $n=0$ and the $n=1$ Landau levels in bilayer graphene are robust against bond disorders 
as long as they are correlated over a few lattice constants.   
We analyse the result in terms of the chiral symmetry for the 
bilayer system.  
Unexpectedly, it is further found that, even in the presence of a potential difference (an electric field) between the top and the bottom layers,
the $n=0$ Landau levels  remain robust although they are 
shifted away from zero energy, while the $n=1$ Landau levels 
exhibit the usual behavior.  
This phenomenon is also discussed in terms of  the effective theory at K and K' points.

\begin{figure}
\includegraphics[scale=0.22]{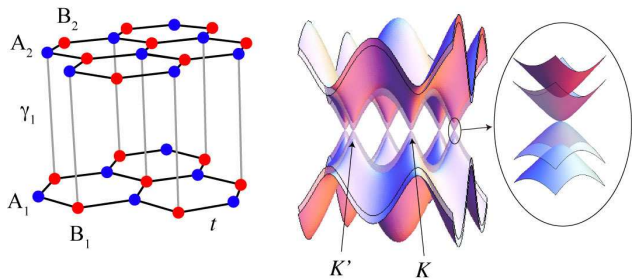}
\caption{(Color Online) For a bilayer graphene we show 
the A-B stacking with the interlayer coupling $\gamma_1$ 
(left panel), and the band 
dispersions (right).  
${\bm K}=(2\pi/3\sqrt{3},2\pi/3)a^{-1}$ and ${\bm K'}=(-2\pi/3\sqrt{3},2\pi/3)a^{-1}$ 
are the corners of the Brillouin zone.
\label{fig1}
}

\end{figure}

{\it Lattice model ---}
In order to investigate the robustness of the zero modes against disorder in bilayer graphene, 
we adopt the following tight-binding lattice model with the Bernal (A-B) stacking. \cite{MF}
We assume that each layer 
can be described by the simple honeycomb lattice, while 
the interlayer coupling, 
$\gamma_1$, connects a site (${\rm B}_1$) on the B sublattice of the bottom layer and a site (${\rm A}_2$) on the A sublattice of the 
top layer just above ${\rm B}_1$(Fig.\ref{fig1}, left). This simplest model accounts for the parabolic dispersion with zero 
gap at K and K' points of bilayer graphene (Fig.\ref{fig1}, right). 
For the randomness, we consider a bond disorder that is spatially 
correlated.  This is  
described by a random component, $\delta t(\bm{r})$, for the 
hopping amplitude in each layers as 
$t({\bm r}) = t + \delta t ({\bm r})$, that is gaussian-distributed with 
a variance $\sigma$ and is correlated in space with a correlation length $\eta$ as 
$\langle \delta t ({\bm r}) \delta t ({\bm r'}) \rangle =
 \langle \delta t^2\rangle \exp(-|{\bm r}-{\bm r}'|^2/4\eta^2)$. 
It is to be noted that in this tight-binding lattice model 
the chiral symmetry is 
exactly preserved even in the presence of disordered components in hopping amplitudes $t$ and $\gamma_1$,\cite{KHA,KHMA} which is 
due to the bipartite structure of the lattice.   
Spin degrees of freedom are suppressed for simplicity.

A randomness is expected to be induced in the hopping amplitudes as a consequence of ripples\cite{Neto} in a monolayer graphene.  
In the case of bilayer, the disorder should be correlated between the two layers if the two layers have a common ripple.  
In present paper, however, we also examine the case where the disorder in two layers are 
uncorrelated  to clarify the generality of the topological protection of the  Landau levels in bilayer graphene.   

The effect of the magnetic field is taken into account by the Peierls substitution $t \to t e^{-2\pi i \theta(\bm{r})}$, such that
the summation of the phases along a loop is equal to the magnetic flux enclosed by the loop in units of the flux quantum $\phi_0=(h/e)$.
The nearest-neighbor distance of the honeycomb lattice 
 is denoted by $a$, while 
the external uniform magnetic flux enclosed by the hexagon of the honeycomb lattice by $\phi$.

\begin{figure}
\includegraphics[scale=0.4]{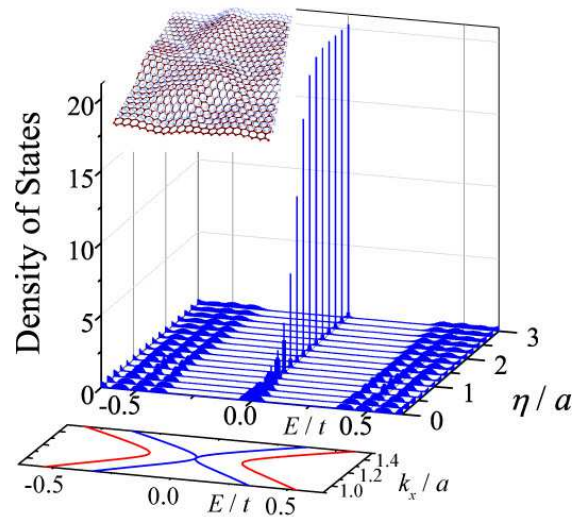}
\caption{(Color Online) Density of states of bilayer graphene 
in a magnetic field plotted for varied correlation length, $\eta$, of 
the bond disorder.   The result is for a system with $10^6$ sites, 
a magnetic field $\phi/\phi_0 = 1/50$, the degree of 
disorder $\sigma /t = 0.115$, inter-layer transfer $\gamma_1 /t = 0.2$, and $\varepsilon /t = 0.0006$.
Energy 
dispersion in zero magnetic field around the K point is also shown with the same energy scale.
Inset: Schematic figure of  correlated ripple.
\label{fig2}
}
\end{figure}

\begin{figure}
\includegraphics[scale=0.32]{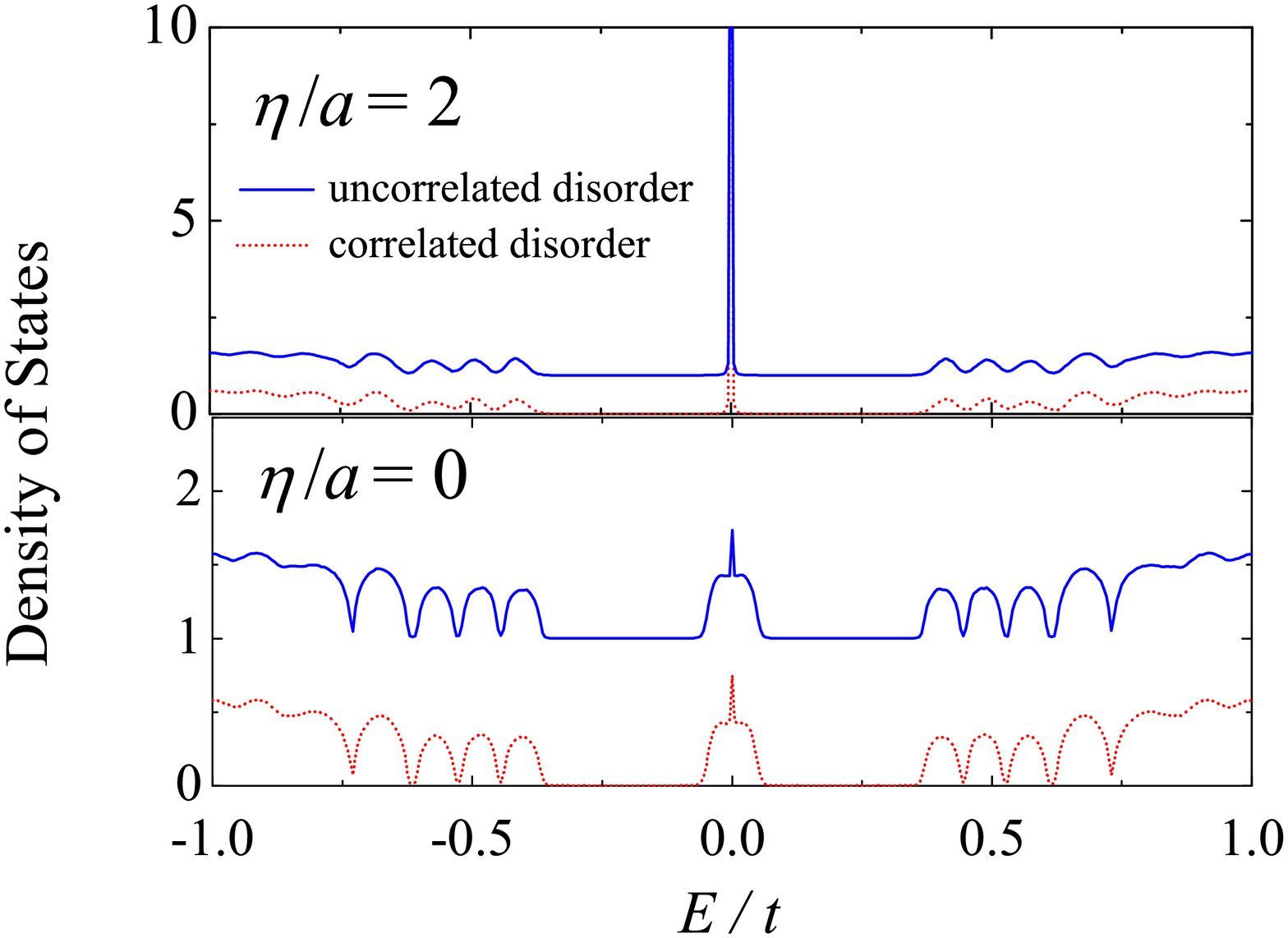}
\caption{(Color Online) Density of states of bilayer graphene for the case of 
correlated disorder {\it between two layers} (dotted curves) and that for uncorrelated disorder  (solid curve)
are shown with an offset 1.0 for uncorrelated disorder. 
The other parameters are the same as in Fig.\ref{fig2}.
\label{fig3}
}
\end{figure}

{\it Zero-mode Landau level ---}
Let us first discuss the robustness of the zero-energy Landau levels.
The density of states $\rho(E) = -\langle{\rm Im}G_{ii}(E+i\varepsilon)/\pi\rangle_i$  with
$G_{ii}(E +i\varepsilon)= \langle i|(E-H+i\varepsilon)^{-1}|i\rangle$ is evaluated by the 
Green function method.\cite{SKM} 
Figure \ref{fig2} displays the density of states for the case where the disorders in the two layers are perfectly correlated.   
We find 
that the zero energy Landau 
level becomes anomalously sharp as soon as the spatial correlation length $\eta$ exceeds 
a few nearest-neighbor distances $a$, which is the same behavior 
as in the case of the monolayer graphene. 
This means that both of the $n=0$ and the $n=1$ Landau levels
at zero energy remain delta-function like in the presence 
of finite-range bond disorder, which confirms  
the prediction based on the effective Hamiltonian at low energies. \cite{KP}    
We also examine the case where the disorder in two layers is 
uncorrelated to find that 
the density of states coincides with those in the correlated case, 
which reveals an insensitivity  to the disorder correlation between two layers (Fig.\ref{fig3}). 


\begin{figure}
\includegraphics[scale=0.49]{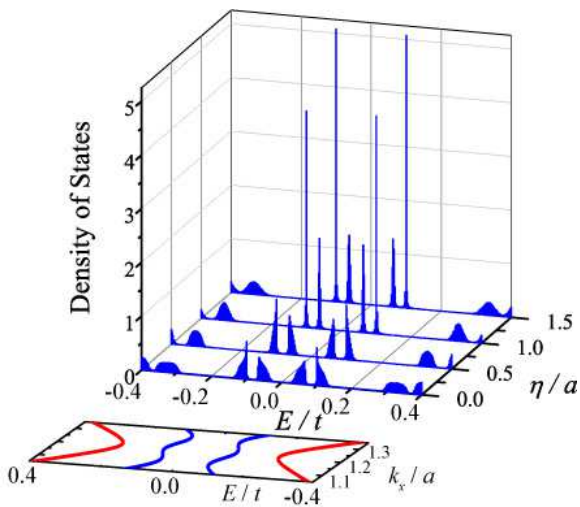}
\caption{(Color Online) Density of states of bilayer graphene in the presence of a 
potential difference $ \Delta /t = 0.1$. The other parameters are the same 
as in Fig. \ref{fig1}.  
Energy 
dispersion in zero magnetic field around the K point is also shown in the same energy scale.
\label{fig4}
}
\end{figure}

\begin{figure}
\includegraphics[scale=0.33]{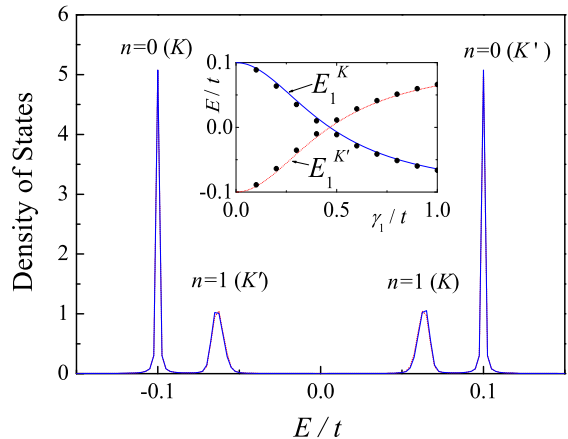}
\caption{(Color Online) Four Landau levels  split by the electric field around zero energy of bilayer graphene when the disorder 
correlation length in each layer is $\eta /a =2$. Here $\Delta /t = 0.1$ and the other parameters are the same as in  Fig. \ref{fig2}.
The results (solid curve) for the case where the disorder is uncorrelated between two layers is plotted on
those(dotted curve) for the case of the same disorder in two layers.  Two curves almost coincide, 
demonstrating clearly the insensitivity to whether 
the disorder is correlated between the layers or not.  
A small width of the $n=0$ levels is an artifact of a finite imaginary energy $\varepsilon$ in the Green function.  
Inset:  The energies of the $n=1$ Landau levels in the presence of disorder are plotted by solid circles as a function of the 
interlayer coupling $\gamma_1$. 
The solid(dotted) curve represents
the perturbational results (Eq. (\ref{E_1})) for K(K') .
\label{fig5}
}
\end{figure}

{\it Electric field effect ---}
In a bilayer graphene we can introduce an energy gap by applying an electric field
perpendicular to the graphene sheet (Fig.\ref{fig4}, inset).\cite{Neto}
While the electric field obviously breaks the inversion symmetry of the system, we should note that it also breaks the chiral symmetry for the effective Hamiltonian around K and K'.  
To examine what happens to the robustness of the Landau levels 
around $E=0$ in such a case, 
we evaluate  the density of states when the potential difference $2\Delta$ is introduced 
between the two layers, where the four-fold degenerated zero-energy Landau levels 
split into four.\cite{MF}   
The present  numerical result, displayed in Fig.\ref{fig4}, shows that, among the split four Landau levels,
the $n=0$ Landau levels that are 
located at energies $E = \pm \Delta$ again become anomalously 
sharp as soon as the disorder is correlated over  few 
lattice constants, 
while the $n=1$ Landau levels are broadened by the disorder 
despite the fact that they are located closer to $E=0$.  
An interesting observation is that 
the energies of these anomalously sharp Landau levels can be tuned by the 
electric field.  We can also note that the $n=1$ Landau levels, 
while not as sharp as the $n=0$ Landau levels, are significantly sharper than higher Landau levels. 
We have also confirmed that the results are insensitive to 
whether the disorder is correlated between the two layers or not (Fig.\ref{fig5}).

{\it Effective theory ---}
The effective Hamiltonian, acting on the envelop functions $(\Psi_{{\rm A}_2}^{\rm K},\Psi_{{\rm B}_1}^{\rm K},\Psi_{{\rm A}_1}^{\rm K}, \Psi_{{\rm B}_2}^{\rm K})$ 
at K-point, 
for bilayer graphene in a perpendicular electric field is given by
\cite{MF,KNA,NKA}
$$
 H = 
\left(\begin{array}{cccc}
\Delta  & \gamma_1&0&v_F\pi_2^\dagger\\
\gamma_1 & -\Delta&v_F\pi_1&0 \\
0 & v_F\pi_1^\dagger &-\Delta &0\\
v_F\pi_2  & 0 &0&\Delta \\
\end{array}
\right),
$$
where $\pi_{\ell} = p_{x\ell}-ip_{y\ell}$ with $\bm{p}_{\ell} = -i\hbar \bm{\partial} +e\bm{A}_{\ell}(e>0)$, and $v_F$ 
the Fermi velocity of a monolayer graphene. 
Here $\ell (=1,2)$ labels the two layers, and 
the gauge field $\bm{A}_{\ell}$ represents the effective gauge field in 
each layer, which  includes the random gauge field 
induced by the random hopping as well as the contribution 
by the external (uniform) magnetic field.  
The operator $\pi_{\ell}$ 
satisfies a commutation relation, $[\pi_{\ell}, \pi_{\ell}^\dagger]=2\hbar e B_{\ell}$, 
with $B_{\ell}=(\nabla \times \bm{A}_{\ell})_z$.  
 Note that the gauge fields  $\bm{A}_1$ and $\bm{A}_2$, and 
consequently the effective magnetic fields $B_1$ and $B_2$  at 
K point, can be  different when the random bonds are 
different between the two layers. 

An important point is that the random component in the gauge field induced by the bond disorder gives rise to 
effective magnetic fields that have opposite signs between 
K and K' points.\cite{Neto}
Still, the effective Hamiltonian is chiral-symmetric, since it satisfies $\Gamma H \Gamma = -H$ as long as $\Delta =0$, where 
the chiral operator $\Gamma$ is given, for a bilayer, in terms of a Pauli matrix $\sigma_z$ as 
$$
 \Gamma = \left(\begin{array}{cc} 
                             \sigma_z & 0 \\
                              0 & \sigma_z 
                              \end{array}\right).
$$
The zero modes for the valley K with no electric field ($\Delta=0$) are, in analogy with the case without disorder,\cite{KNA}  given by
$$
 \bm{\Psi}_{n=0}^{\rm K} = \mvecfour{0}{0}{\psi_{0}^{(1)}}{0}\quad  {\rm and} \quad \bm{\Psi}_{n=1}^{\rm K}
 = \mvecfour{\psi_{0}^{(2)}}{0}{-\frac{\gamma_1}{v_F}\psi_{1}}{0}
$$
with $\pi_1 \psi_{0}^{(1)} = 0$, $\pi_1 \psi_1 = \psi_{0}^{(2)}$ and $\pi_2 \psi_{0}^{(2)}=0$.
Note that these zero modes are also eigenstates of the chiral operator $\Gamma$.

Following the argument by Aharonov and Casher\cite{AC}, we adopt the Coulomb gauge $\partial_x^2 A_x + \partial_y^2 A_y =0$, and  
express the gauge field as  
$\bm{A}=(-\partial_y \varphi, \partial_x \varphi)$. The operator $\pi$ 
is then expressed as 
$\pi_{\ell}= -2i\hbar [\partial_{z^*} + {2\pi}(\partial_{z^*}\varphi_{\ell}) /\phi_0]$
with $z \equiv (x+iy)/2$. The solution to $\pi_\ell \psi_{0}^{(\ell)} = 0$ with $\ell=1,2$ is given by
$ \psi_{0}^{(\ell)} = f_{\ell}(z) \exp(-2\pi \varphi_{\ell}/\phi_0)$, where
$f_{\ell}(z)$  a polynomial in $z$.\cite{AC}  In general, 
the solution $\psi_1$ takes the form
$
 \psi_1 = f_2(z) \exp(-2\pi \varphi_1/\phi_0) F(z,z^*), 
$
with $\partial_{z^*} F(z,z^*) = \exp(2\pi(\varphi_1 -\varphi_2)/\phi_0)$.
If the bond disorders in two layers are the same ($\varphi_1 = \varphi_2$), 
the function $F(z,z^*)$ is reduced to $z^*$.\cite{Kail}
It is straightforward to apply these arguments to the effective Hamiltonian for the valley K',\cite{MF,KNA,NKA} which implies that 
the four-fold degenerated zero-energy Landau levels exist irrespective of 
presence or absence of the disorder in gauge fields.

When the electric field is switched on ($\Delta \neq 0$), the above chiral symmetry is broken. 
We can still show, however, that the state $\bm{\Psi}_{n=0}^{\rm K}$ remains to be an exact eigenstate of the Hamiltonian with 
the eigenvalue $\epsilon_0^{\rm K}=-\Delta$, so that {\it the broadening due to disorder is absent} as in the case of $\Delta=0$ is demonstrated. 
The state $\bm{\Psi}_{n=1}^{\rm K}$, on the other hand, is not an exact eigenstate for $\Delta \neq 0$. 
It is therefore natural to expect that the broadening occurs for the 
Landau levels corresponding to $\bm{\Psi}_{n=1}^{\rm K}$, as is actually seen in our numerical results (Figs. \ref{fig4} and \ref{fig5}). 
We can also note that even for such states, the broadening itself 
is likely to be significantly smaller than those for higher Landau levels, 
which comes from the anomalous character of 
the unperturbed Landau level $\bm{\Psi}_{n=1}^{\rm K}$ (Fig. \ref{fig4}).

The eigenvalue $E_1^{\rm K}$ for the state $\bm{\Psi}_{n=1}^{\rm K}$ can be estimated, in the absence of disorder, from the perturbation 
with respect to $\Delta$, which gives 
\begin{equation}
 E_1^{\rm K} = \bigg( 1 - \frac{2\gamma_1^2}{2v_F^2\hbar e B+\gamma_1^2} \bigg)\Delta,
 \label{E_1}
\end{equation}
where $B=B_1 = B_2$ denotes the uniform external magnetic field.
Relations to the tight-binding parameters are given by
$v_F = (3/2)at/\hbar$ and $\phi = (3\sqrt{3}/2)Ba^2$. 
The energy $E_1^{\rm K} $ is then estimated as $E_1^{\rm K} \simeq  0.7\Delta$ for $\phi = (1/50)\phi_0$ and $\gamma_1/t =0.2$, 
which accurately agrees with the present 
numerical result (Fig.\ref{fig5}).  
The same argument for the effective Hamiltonian at K' point
leads to the Landau levels at $E_0^{\rm K'} = -E_0^{\rm K}=\Delta$ and 
$E_1^{\rm K'} =-E_1^{\rm K} \simeq -0.7 \Delta$. 
We also examine the positions of the $n=1$ Landau levels for various values of the interlayer coupling $\gamma_1$ to 
confirm that their positions are in good agreement with the perturbational result (Eq.(\ref{E_1})) for the range $0< \gamma_1< t$ (Fig.\ref{fig5}, inset). 

The peak heights of these valley-split Landau levels depend on 
the effective magnetic field for each valley. 
Since the effective fields induced by ripples  
in K and K' points have the opposite sign 
with the same magnitude,\cite{Neto} the degeneracies of 
the Landau levels can be different for K and K', 
although their sum should be a constant.\cite{Kail} 
In our numerical results, however, no significant difference in the peak heights for the valley-split 
Landau levels is seen (Figs.\ref{fig4} and \ref{fig5}). This can be attributed to the fact that the 
present density of states is an average over the sample. Our sample-size 
is much larger than the correlation length of bond disorder and the periodic boundary 
condition is adopted along the strip geometry.\cite{KHA} It is  therefore  expected that 
the local fluctuation of the magnetic field due to the bond disorder (ripples) is 
canceled.

{\it Conclusions ---} 
We have demonstrated, both numerically and analytically, that the zero-energy Landau levels of bilayer graphene become 
anomalously sharp when the bond disorder is correlated over a few lattice constants. The anomaly 
is shown to be insensitive to the disorder correlation between the top and the bottom layers, which suggests a 
relevance of the chiral symmetry to the present anomaly as in the case of the monolayer graphene.  
Another new finding is that  the anomaly at the $n=0$  Landau level 
persists 
even in the case where the chiral symmetry for each valley is broken by the potential difference between two layers.
The splitting of the pair of sharp Landau levels is controlled by the electric field, 
and their peak heights reflect the effective magnetic field strength at each valley. 
The anomalous sharpness of these levels found here 
may help to detect experimentally the local fluctuation of 
the effective magnetic field arising from ripples as unbalanced peak heights of these Landau levels in the {\it local} density of states.

\begin{acknowledgments}
We wish to thank Yoshiyuki Ono and Takahiro Morimoto
for useful discussions and comments.
The work was supported in part by Grants-in-Aid for Scientific Research,
Nos. 22540336 and 23340112  from JSPS. YH was also supported in part by Grants-in-Aid No.23654128 (JSPS),
No.  22014002 (MEXT) and National Science Foundation under Grant No. PHY05-51164.

\end{acknowledgments}

%\bibliography{apssamp.bib}

\begin{thebibliography}{10}
%
\bibitem{Geim} K.S. Novoselov et al, 
Nature {\bf 438}, 197 (2005).
\bibitem{Kim} Y. Zhang, Y.W. Tan, H.L. Stormer, and P. Kim, 
Nature {\bf 438}, 201 (2005).
\bibitem{OGM} P.M. Ostrovsky, I.V. Gornyi, and A.D. Mirlin, Phys. Rev. B {\bf 77}, 195430 (2008).
\bibitem{Guinea} F. Guinea, B. Horovitz, and P. Le Doussal, Phys. Rev. B 
{\bf 77}, 205421 (2008).
\bibitem{LFSG} A.W.W. Ludwig et al, 
Phys. Rev. B {\bf 50}, 7526 (1994).
\bibitem{Hatsu10} Y. Hatsugai, arXiv:1008.4653.
\bibitem{KHA} T. Kawarabayashi, Y. Hatsugai, and H. Aoki, Phys. Rev. Lett. {\bf 103}, 156804 (2009); Physica E {\bf 42}, 759 (2010).
\bibitem{KMHA} T. Kawarabayashi, T. Morimoto, Y. Hatsugai, and H. Aoki, Phys. Rev. B {\bf 82}, 195426 (2010).
\bibitem{GZKPMM} A.J.M. Giesbers, U. Zeitler, M.I. Katsnelson, L.A. Ponomarenko, T.M. Mohiuddin, and J.C. Maan, Phys. 
Rev. Lett. 99, 206803 (2007).
\bibitem{KHMA} T. Kawarabayashi, Y. Hatsugai, T. Morimoto, and H. Aoki, Phys. Rev. B {\bf 83}, 153414 (2011).
\bibitem{KKS} S. Katayama, A. Kobayashi, and Y. Suzuura, J. Phys. Soc. Jpn. {\bf 75}, 054705 (2006).
\bibitem{GFMP} M.O. Goerbig, J.-N. Fuchs, G. Montambaux, and F. Pi\'{e}chon, Phys. Rev. B {\bf 78}, 045415 (2008).
\bibitem{TSKNK} N. Tajima, S. Sugawara, R. Kato, Y. Nishio, and K. Kajita, Phys. Rev. Lett. {\bf 102}, 176403 (2009).
\bibitem{AC} Y. Aharonov and A. Casher, Phys. Rev. A {\bf 19}, 2461 (1979).
\bibitem{Nakahara} M. Nakahara, {\it Geometry, Topology, and Physics}, 2nd ed. (Taylor \& Francis, 2003).
\bibitem{MF} E. McCann and V.I. Falko, Phys. Rev. Lett. {\bf 96}, 086805 (2006). 
\bibitem{KP} M.I. Katsnelson and M.F. Prokhorova, Phys. Rev. B {\bf 77}, 205424 (2008).
\bibitem{Kail} J. Kailasvuori, Europhys. Lett. {\bf 87}, 47008 (2009).
\bibitem{Weitz}R. T. Weitz, M. T. Allen, B. E. Feldman, J. Martin, and 
A. Yacoby, Science {\bf 330}, 812 (2010).
\bibitem{Martin} J. Martin, B. E. Feldman, R. T. Weitz, M. T. Allen, and A. Yacoby, Phys. Rev. Lett. {\bf 105}, 256806 (2010).
\bibitem{OBSHR} T. Ohta, A. Bostwick, T. Seyller, K. Horn, and E. Rotenberg, Science {\bf 313}, 951 (2006).
\bibitem{Castro} E.V. Castro et al., Phys. Rev. Lett. {\bf 99}, 216802 (2007).
\bibitem{Neto} A.H. Castro Neto et al, Rev. Mod. Phys. {\bf 81}, 109 (2009).
\bibitem{SKM} L. Schweitzer, B. Kramer, and A. MacKinnon, J. Phys. C {\bf 17}, 4111 (1984).

\bibitem{KNA} M. Koshino, T. Nakanishi, and T. Ando, Phys. Rev. B {\bf 82}, 205436 (2010). 
\bibitem{NKA}T. Nakanishi, M. Koshino, and T. Ando, Phys. Rev. B {\bf 82}, 125428 (2010). 
\end{thebibliography}

\vfill
\end{document}